\documentclass{mn2e}
\usepackage{psfig}
\usepackage{epsf}
\newcommand{\ltaraw}{$\; \buildrel < \over \sim \;$}
\newcommand{\lta}{\lower.5ex\hbox{\ltaraw}}
\newcommand{\gtaraw}{$\; \buildrel > \over \sim \;$}
\newcommand{\gta}{\lower.5ex\hbox{\gtaraw}}

\loadboldmathitalic 
\title [${\rm H_o}$ and quintessence cosmologies]
{An investigation of gravitational lens determinations of ${\rm H_o}$ in 
quintessence cosmologies}
\author[G. F. Lewis \& R. A. Ibata]
{Geraint F. Lewis$^{1}$ \& Rodrigo A. Ibata$^{2}$\\
$^{1}$
Anglo-Australian Observatory, P.O. Box 296, Epping, NSW 1710, Australia:
Email \tt{gfl@aaoepp.aao.gov.au}\\
$^{2}$
Observatoire de Strabourg, 11, rue de l'Universite, F-67000, Strasbourg, 
France:
Email \tt{ibata@astro.u-strasbg.fr}
}
\date{\today}
\begin{document} 
\maketitle 
\begin{abstract}
There is growing  evidence that the majority of  the energy density of
the universe is  not baryonic or dark matter, rather  it resides in an
exotic  component   with  negative  pressure.   The   nature  of  this
`quintessence' influences our view  of the universe, modifying angular
diameter and luminosity distances. Here, we examine the influence of a
quintessence component  upon gravitational lens time  delays.  As well
as a static  quintessence component, an evolving equation  of state is
also  considered.  It  is  found that  the  equation of  state of  the
quintessence component  and its evolution  influence the value  of the
Hubble's  constant derived  from gravitational  lenses.   However, the
differences   between  evolving   and  non-evolving   cosmologies  are
relatively small.  We undertake a  suite of Monte Carlo simulations to
examine the potential constraints that  can be placed on the universal
equation of  state from the  monitoring of gravitational  lens system,
and demonstrate that  at least an order of  magnitude more lenses than
currently known will have to  be discovered and analysed to accurately
probe any quintessence component.
\end{abstract}
\begin{keywords} 
cosmology: theory -- cosmological parameters -- gravitational lensing
\end{keywords} 

\section{Introduction}\label{introduction}
The searches for supernovae at cosmological distances have proved very
successful,  providing evidence  that, while  topologically  flat, the
majority  of energy  in  the Universe  is  in the  form  of an  exotic
component with  negative pressure (Riess  et al.  1999;  Perlmutter et
al.   1999).  The  recent  identification of  a  supernova at  $z=1.7$
(Riess et  al.  2001)~\footnote{It should be noted,  however, that the
influence  of gravitational  lensing  on SN1997ff  needs  to be  fully
addressed before  its true cosmological significance  can be addressed
(Lewis  \&  Ibata 2001;  Moertsell,  Gunnarsson  \& Goobar~2001)}  has
provided further weight to these  claims (Turner \& Riess 2001), which
suggest that this component may differ from the classical cosmological
constant $\Lambda$.  Termed `quintessence', or more colloquially `dark
energy', this has an equation of state of the form $P = w \rho$, where
$P$  is the pressure  and $\rho$  the density.   A $w  < -\frac{1}{3}$
opposes the action of gravity and drives the cosmological expansion to
accelerate.  Linder (1988a; 1988b) has examined the physical nature of
various    quintessence   components;    with   $w=0$    equating   to
non-relativistic  matter (dust),  $w=\frac{1}{3}$ being  radiation and
$w=-1$,  a classical  cosmological constant.   More  exotic components
are;   massless   scalar   fields   $w=1$,  cosmic   string   networks
$w=-\frac{1}{3}$,    and     two-dimensional    topological    defects
$w=-\frac{2}{3}$.    As  well   as  the   supernova   programs,  other
approaches,  such  as  gravitational  lensing  statistics  (Cooray  \&
Huterer  1999), geometrical  probes  of the  ${\rm Ly_\alpha}$  forest
(Hui, Stebbins  \& Burles 1999) and galaxy  distributions (Yamamoto \&
Nishioka  2001), and  classical angular-size  redshift tests  (Lima \&
Alcaniz  2001), will  provide  complementary probes  of the  universal
equation of state.

The value of  the quintessence component, $w$, influences  our view of
the  universe, modifying  the various  distances used  in  mapping the
cosmos.   This paper  concerns itself  with  the influence  of $w$  on
angular   diameter   distances,   especially   in  relation   to   the
determination of  the Hubble's constant  from the measurement  of time
delays in gravitational lens  systems.  Unlike local determinations of
Hubble's constant (e.g. Freedman et al. 2001), the cosmological nature
of  gravitational lenses  means that  they are  more sensitive  to the
underlying cosmological  parameters.  Section~\ref{background} briefly
covers the  basic formulae for generalized  angular diameter distances
in  quintessence  cosmologies,  while  in  Section~\ref{timedelay}  we
consider  the influence  of $w$  on  the determination  of $H_o$  from
lensed systems.  Section~\ref{evolve}  extends this analysis to simple
models     of    an     evolving    quintessence     component.     In
Section~\ref{montecarlo}  a  series  of  Monte Carlo  simulations  are
undertaken to  estimate the efficacy  of this approach in  probing the
cosmological  equation of state,  while in  Section~\ref{speculate} we
speculate   on   the   possability   that  current   observations   of
gravitational lens  systems may suggest that  $w<-1$.  The conclusions
of this study are presented in Section~\ref{conclusions}.

\begin{figure*}
\centerline{ \psfig{figure=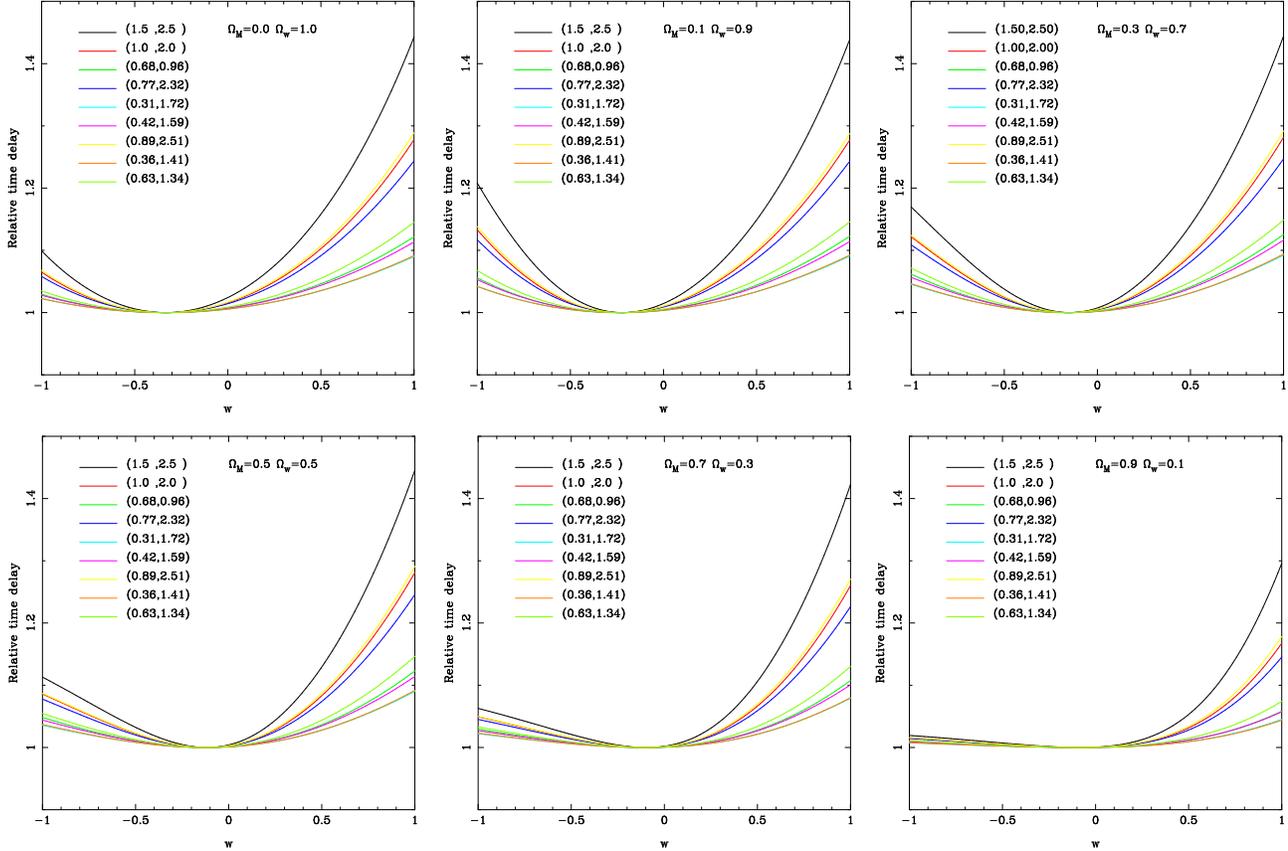,height=14.5cm,angle=270.0} }
\caption{The relative  time delays for  several flat cosmologies  as a
function of  the equation  of state,  $w$.  Each curve  is is  for the
lowest value of  the time delay (the redshift  of the lens-source pair
is given in brackets).  Each panel presents a different combination of
$\Omega_m$ and  $\Omega_w$; note that $w=0$ corresponds  to a universe
composed entirely of matter in all cases.}
\label{fig1}
\end{figure*}

\section{Generalized Angular Diameter Distances}\label{background}
While  there has  been  a resurgence  in  quintessence cosmology,  the
generalized cosmological  equations for such  universes were presented
more  than   a  decade  ago  by  Linder   (1988a;1988b),  including  a
generalized form  of the Dyer-Roeder  equation for the evolution  of a
bundle of rays traveling from  a distant source (Dyer \& Roeder 1973).
Expressing  the angular  diameter distance  as  $D =  (c/H_o) r$,  the
generalized beam equation is given by
\begin{equation}
\ddot{r} + \left[\frac{3+q(z)}{1+z}\right]\dot{r} + \sum_w 
\frac{3 r (1+w)\alpha_w(z)\Omega_w(z)}{2(1+z)^2}  = 0 
\label{linder}
\end{equation}
where $\Omega_w(z)$ is  the density, in units of  the critical density
$\rho_c(z)$,  of a  contributor  to the  total  energy-density of  the
universe with an  equation of state $w  = P / \rho$, and where  $P$ is its
pressure and $\rho$ is its density. This is given by
\begin{equation}
\Omega_w(z) \equiv \frac{\rho_w(z)}{\rho_c(z)} = 
\Omega_w(0)(1+z)^{3(1+w)}\left[\frac{H_o}{H(z)}\right]^2 
\label{Omega}
\end{equation}
where  $\Omega_w(0)$ is  the  contribution of  this  component to  the
present energy-density  budget at the present epoch,  and the critical
density is given by $\rho_c(z) = 3 H(z)^2 / 8\pi G$. Here, $H(z)$ is a
generalized form of the Hubble parameter and is given by;
\begin{equation}
H(z) = H_o\left[ \sum_w \Omega_w(0)(1+z)^{3(1+w)} - K(1+z)^2\right]^\frac{1}{2}
\label{hubble}
\end{equation}
where  $K = \Omega_o  - 1$,  and $\Omega_o  \equiv \Omega(0)  = \sum_w
\Omega_w(0)$,  is related to  the overall  curvature of  the universe.
$q(z)$  is a  generalized form  of the  deceleration parameter  and is
given by
\begin{equation}
q(z) = \frac{1}{2} \frac{\sum_w\Omega_w(0)(1+3w)(1+z)^{1+3w}}
{\sum_w\Omega_w(0)(1+z)^{1+3w} - K }
\label{decell}
\end{equation}
Finally,    Equation~\ref{linder}   also   contains    the   parameter
$\alpha_w(z)$ which represents how much  of the fluid lies in the beam
and  influences the the  evolution of  a ray  bundle.  For  a universe
containing   matter,  the   solution  to   Equation~\ref{linder}  with
$\alpha_0(z)=1$  represents  the  classic  Dyer-Roeder  `filled  beam'
distance, while $\alpha_0(z)=0$ is  the `empty beam' distance (Dyer \&
Roeder 1973).

\begin{figure*}
\centerline{ \psfig{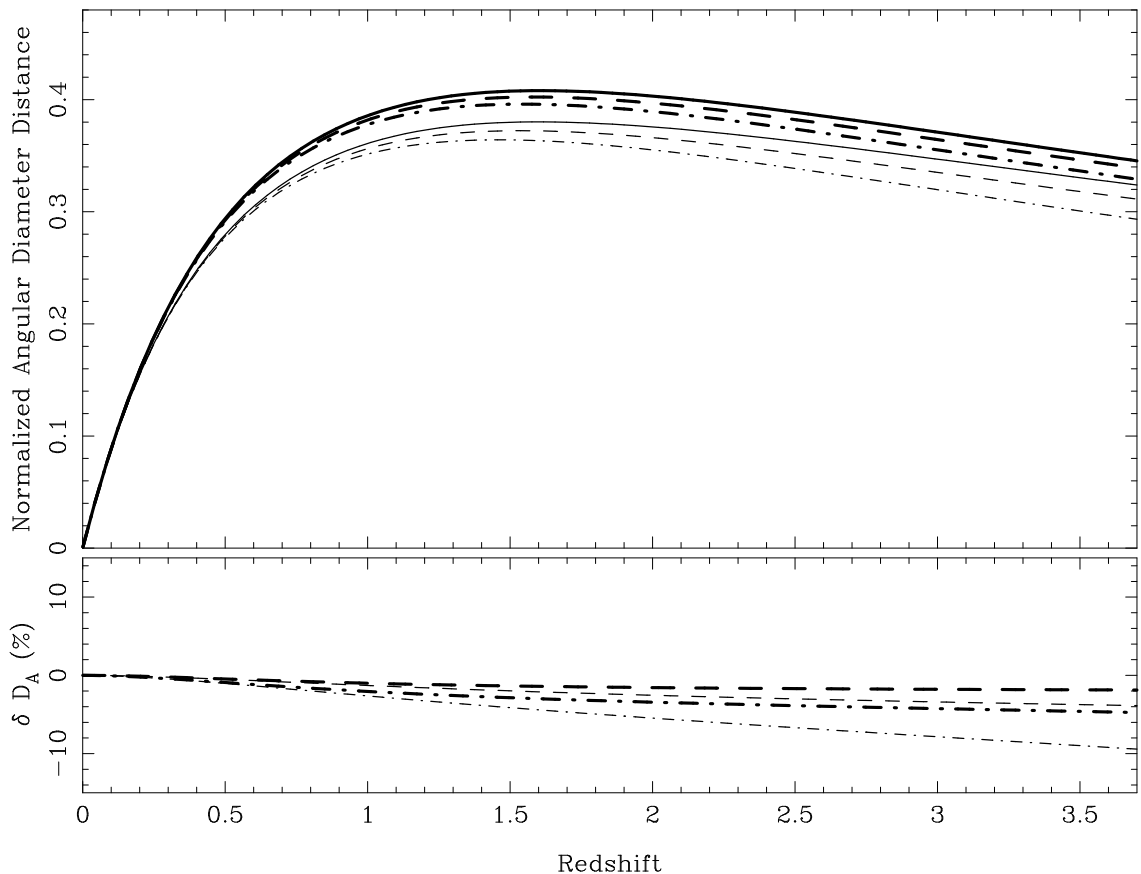} }
\caption{The  top panel presents  the angular  diameter distance  of a
source  at $z$  for an  observer at  $z=0$, for  the  evolutionary $w$
models discussed in the text.  The thick lines correspond to universes
with $w_0=-1$,  while the thin line represents  $w_0=-0.7$.  The solid
line represents  a non-evolving model  with $w_1=0$, while  the dashed
lines evolve as $w_1=0.2$ and  the dot-dashed lines as $w_1=0.4$.  The
lower panel  presents the  percentage difference between  the evolving
models and non-evolving model.}
\label{fig2}
\end{figure*}

When solving Equation~\ref{linder}, the boundary conditions need to be
defined. These are
\begin{eqnarray}
r(z_0,z_0) = 0, \\
\left. \frac{dr(z_0,z)}{dz} \right|_{z=z_0} 
= (1+z_0)^{-1}\left[\frac{H_o}{H(z)}\right]
\label{boundary}
\end{eqnarray}
Equation~\ref{linder} was  integrated using a  Runge-Kutta scheme (the
{\tt rksuite} package from  {\tt www.netlib.org}) and compared to both
analytic  results   and  the  minimum  angular   extent  redshifts  in
quintessence  cosmologies  as tabulated  in  Lima  \& Alcaniz  (2000);
excellent  agreement  was found.   Throughout  this work,  filled-beam
distances $\alpha_w(z)=1$ are employed.

\begin{figure*}
\centerline{ \psfig{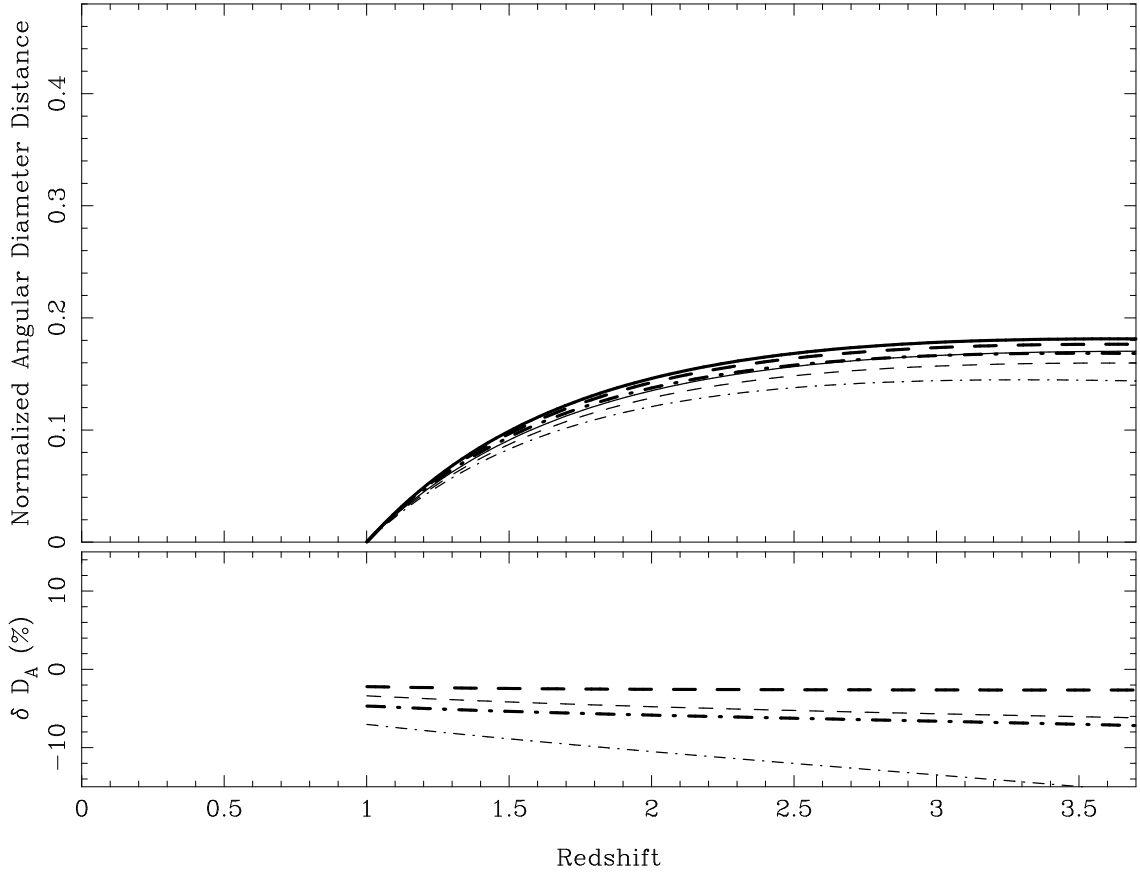} }
\caption{As for Figure~\ref{fig2}, but for an observer at $z=1.0$.}
\label{fig4}
\end{figure*}

\section{Gravitational lens time delays}\label{timedelay}
Refsdal  (1964) was  the first  to note  that  cosmological parameters
could be determined  from the measurement of a  time delay between the
relative  paths taken  by light  though a  gravitational  lens system.
Since the  discovery of multiply-imaged  quasars, this has  become the
goal of  a number  of monitoring campaigns  (e.g.  Cohen et  al. 2000;
Oscoz et al. 2001; Patnaik \& Narashima 2001), although these analyses
are  frustrated  by degeneracies  in  the  derived  mass models.   The
cosmological  model  simply   enters  the  determination  of  Hubble's
constant;
\begin{equation}
H_o \Delta t \propto \frac{ r_{ol} r_{os} }{r_{ls} }
\label{hubbletime}
\end{equation}
where $r_{ij}$  are the normalized angular  diameter distances between
and observer $(o)$, lens $(l)$ and source $(s)$, and $\Delta t$ is the
measured time delay between an image pair.

Giovi  \& Amendola  (2001) examined  the influence  on  a quintessence
component on  gravitational lens time delays and  the determination of
Hubble's constant. Their analysis,  however, was mainly concerned with
the influence of the clumping  of material and the dependence on $H_o$
of whether  distances are empty-beam  or full-beam. Here,  a different
approach     is     considered;     Solving     Equation~\ref{linder},
Equation~\ref{hubbletime}  is evaluated  for a  range  of quintessence
components. For  the study,  the redshifts of  the lens and  source in
seven gravitational  lens systems that  are favourable for  time delay
measures  were  considered  [Table~2  in  Giovi  \&  Amendola  (2001),
Q0957+561  and  B1608+656],  plus   two  fiducial  redshift  pairs  of
$(z_l=1.5,z_s=2.5)$ and $(z_l=1.0,z_s=2.0)$. While there are currently
no  lensed systems with  established time  delays at  these particular
redshifts, there  are several  potential systems; e.g.   the quadruple
lens  H1413+117  at  a  redshift   of  2.55,  with  a  lens  redshift,
established  from  prominent  absorption  features, at  $\sim1.5$  and
HE1104-1805 at  a redshift of 2.31  and an estimated  lens redshift of
$z\sim1$.   It  is  assumed  throughout  that the  universe  is  flat,
$\Omega_m + \Omega_w = 1$.

\begin{table}
\centering
\begin{tabular}{ccc}
$\Omega_m$ & $\Omega_w$    & $w_{min}$  \\ \hline
0.0        & 1.0           & -0.33      \\
0.1        & 0.9           & -0.22      \\
0.3        & 0.7           & -0.15      \\
0.5        & 0.5           & -0.12      \\
0.7        & 0.3           & -0.10      \\
0.9        & 0.1           & -0.09      \\ \hline
\end{tabular}
\caption{\label{table1} The minima for the curves in Figure~\ref{fig1}
for the various cosmologies under consideration.}
\end{table}

Figure~\ref{fig1} presents  the results  of this analysis;  six panels
are  presented, each  for a  different combination  of  $\Omega_m$ and
$\Omega_w$.  A   greyscale-coded  line  for  each   redshift  pair  is
presented.   The abscissa  presents the  equation of  state parameter,
$w$,  while  the  ordinate  presents  the  relative  time-delay;  this
represents  the  change  in  the  time  delay  for  a  fixed  Hubble's
constant. Conversely, this is the relative value of $H_o$ for a system
with a fixed time delay. Each curve is normalized to the minimum value
of the time delay.  The  relative time delay depends quite strongly on
the value  of $w$, with the $(z_l=1.5,z_s=2.5)$  possessing changes of
$\sim40\%$  at  $w=1$  as  compared  to $w\sim0$,  although,  for  the
observed lensing systems,  the same range in $w$  produces a change of
$\sim10\%$  in   the  same  quantity.   Interestingly,   for  a  fixed
combination of $\Omega_m$ and $\Omega_w$, all the curves, irrespective
of the redshift  of the lens and source possess a  minimum at the same
value  of   $w$.  The  location   of  the  minimum  is   tabulated  in
Table~\ref{table1}.

At present, the general  analytic solution to Equation~\ref{linder} is
quite complex (Giovi \& Amendola  2001) and it is difficult to further
analyze the  minimum seen in Figures~\ref{fig1}.  For  one case, where
$\Omega_m = 0.0$ and $\Omega_w = 1.0$, however, analytic solutions for
the relative  angular diameter distance are straight  forward.  As the
overall curvature  is flat, the  angular diameter distance  between us
and a distant source is
\begin{equation}
D_A(z) = \frac{ D_c(z) }{1 + z}
\label{cosmo1}
\end{equation}
where $D_c(z)$ is the comoving distance and is given by
\begin{equation}
D_c(z) = \frac{c}{H_o} \int_o^z \frac{dz'}{E(z')}
\label{cosmo2}
\end{equation}
where $E(z) = \Omega_w(0)  ( 1+z )^{\frac{3}{2}(1+w)}$.  Again, as the
overall cosmology  is flat, the angular diameter  distance between the
source and the lens is given by;
\begin{equation}
D_A(z_1,z_2) = \frac{1}{1+z_2}\left[ D_c(z_2) - D_c(z_1) \right]
\label{flat}
\end{equation}
With this,  the relative  time delays for  differing values of  $w$ is
seen to be:
\begin{equation}
H_o\left[z_1,z_2\right] \propto 
\frac{1}{1+z_1}\frac{2}{1+3w}
\left[\frac{ (1 - (1+z_1)^f)(1 - (1+z_2)^f)}
{(1+z_1)^f - (1+z_2)^f}\right]
\label{mathematica}
\end{equation}
where $f  = -\frac{1}{2}(1+3w)$. This function possesses  a minimum at
$w = -\frac{1}{3}$, which is  independent for the redshift pairs under
consideration, as seen in Figure~\ref{fig1}.

The  preceding  section   has  considered  arbitrary  combinations  of
cosmological parameters. Here, the impact  of the choice of $w$ on the
estimation  of   $H_o$  in   the  favoured  cosmological   model  with
$\Omega_m=0.3$    and    $\Omega_w=0.7$    (top   right    panel    in
Figure~\ref{fig1}) is examined in  more detail.  Often, a cosmological
model where  $\Omega_m=1$ is choosen  when estimating $H_o$  from time
delays, which is equivalent to considering $w=0$ in Figure~\ref{fig1}.
An  accelerating   cosmology,  as  suggested   from  the  cosmological
supernovae programs,  requires $w<-\frac{1}{3}$. It  is interesting to
note that  in this  cosmology, the minimum  $H_o$ occurs  at $w=-0.15$
(Table~1)  the values of  $H_o$ derived  at $w=0$  or $w=-\frac{1}{3}$
will  be  very similar.   Considering  more  negative  values of  $w$,
adopting classical cosmological constant, with $w=-1$ results in $H_o$
values that  are $\sim5\%$  different to the  all matter  case ($w=0$).
For several combinations of  redshifts, however, (including the chosen
fiducial points)  more substatial  differences of $\sim12-15\%$  in the
determined  value  of  $H_o$.   Considering  $w<-1$  results  in  more
significant discrepancies, a point that will be returned to later.

\section{Evolving quintessence}\label{evolve}
One  interesting  aspect of  quintessence  cosmology  is that,  unlike
classical $\Omega_m,\Omega_\Lambda$  cosmologies, it is  possible that
the equation  of state  $w$ may  vary over the  cosmic history  of the
universe.  Typically, a linear model where $w(z) \sim w_0 + w_1 z$ has
been adopted (e.g.  Goliath et al.  2001).  In the following analysis,
it is assumed  that $\Omega_m = 0.3$ and  $\Omega_w = 0.7$, consistent
with the recent supernova  experiments (Riess et al.  1999; Perlmutter
et  al.  1999)  results.  Four  models for  the evolution  of  $w$ are
adopted [taken  from Linder (2001)], namely,  $(w_0,w_1) = (-0.7,0.2),
(-0.7,0.4), (-1.0,0.2)$ and $(-1.0,0.4)$. As the cosmology is flat, we
follow the  approach outlined in  Equations~\ref{cosmo1}, \ref{cosmo2}
and \ref{flat}, with
\begin{equation}
E(z) = \sqrt{ (1+z)^3 \Omega_m + f_w(z) \Omega_w }
\end{equation}
and
\begin{equation}
f_w(z) = \exp\left[ 3 \int_o^z \frac{1 + w(z') }{1 + z'} dz'\right].
\end{equation} 

\begin{table}
 \centering
 \begin{tabular}{cccccc}
 $z_l$ & $z_s$ & $(-1.0,0.2)$ &$(-1.0,0.4)$&$(-0.7,0.2)$&$(-0.7,0.4)$ \\ \hline
1.50 & 2.50 &
0.997 &
1.003 &
1.012 &
1.043 \\
1.00 & 2.00 &
1.000 &
1.005 &
1.010 &
1.029 \\
0.68 & 0.96 &
1.003 &
1.007 &
1.007 &
1.015 \\
0.77 & 2.32 &
1.001 &
1.005 &
1.009 &
1.026 \\
0.31 & 1.72 &
1.002 &
1.005 &
1.005 &
1.012 \\
0.42 & 1.59 &
1.002 &
1.006 &
1.006 &
1.014 \\
0.89 & 2.51 &
1.000 &
1.005 &
1.010 &
1.030 \\
0.36 & 1.41 &
1.002 &
1.006 &
1.005 &
1.012 \\
0.63 & 1.34 &
1.002 &
1.006 &
1.008 &
1.017 \\
 \hline
 \end{tabular}
 \caption{\label{table2} The  ration of the time  delay (or conversely
$H_0$) for  the cosmologies  with linearly evolving  quintessence. The
first two  columns are the lens  and source redshifts,  while the next
four  columns present  the  time delays  for  $(w_0,w_1)$. Those  with
$w_0=-1$ are normalized with respect to $(-1.0,0.0)$, while those
with $w_0=-0.7$ are normalized with respect to $(-0.7,0.0)$.}
 \end{table}

Figures~\ref{fig2}  and   \ref{fig4}  present  the   angular  diameter
distance in these  evolving models for an observer  at $z=0$ and $z=1$
respectively.   It is  interesting to  note  that for  an observer  at
redshifts  greater than  zero  the fractional  difference between  the
angular   diameter  distances   for  an   evolving   and  non-evolving
quintessence component  does not converge to zero  as $z_s \rightarrow
z_l$,  rather, as  seen in  the lower  box of  Figure~\ref{fig4}, they
converge to  a non-zero value.  This  seems somewhat counter-intuitive
given the  convergence of distances  for an observer at  redshift zero
(Figure~\ref{fig2}).  Employing Equation~\ref{flat} and setting $z_2 =
z_1 + \Delta z$, the angular diameter distance becomes;
\begin{equation}
D_A(z_1,z_1+ \Delta z) \sim \frac{\Delta z}{1+z_1} 
\left.\frac{d D_c(z)}{dz}\right|_{z=z_1}
\end{equation}
where $\Delta  z \ll 1+z_1$.   Therefore, the asymptotic ratio  of the
angular diameter distance in a non-evolving cosmology, with a constant
$w_0$, and  one where $w(z) = w_0  + z w_1$, as  $z_2 \rightarrow z_1$
simply becomes;
\begin{equation}
\frac{D_A(z_1,w_0)}{D_A(z_1,w(z))} = 
\sqrt{ 
\frac{
\Omega_m (1+z_1)^3 + \Omega_w (1+z_1)^{3(1+w_0)}
}
{
\Omega_m (1+z_1)^3 + \Omega_w g(z_1) 
} 
} ,
\end{equation}
where
\begin{equation}
g(z) = \exp\left\{ 3 ( z w_1 + ( 1 + w_o - w_1 )\log[1 + z] ) \right\}.
\end{equation}
With  the values  under  consideration in  Figure~\ref{fig4}, this  is
precisely the asymptote seen.

\begin{figure}
\centerline{ \psfig{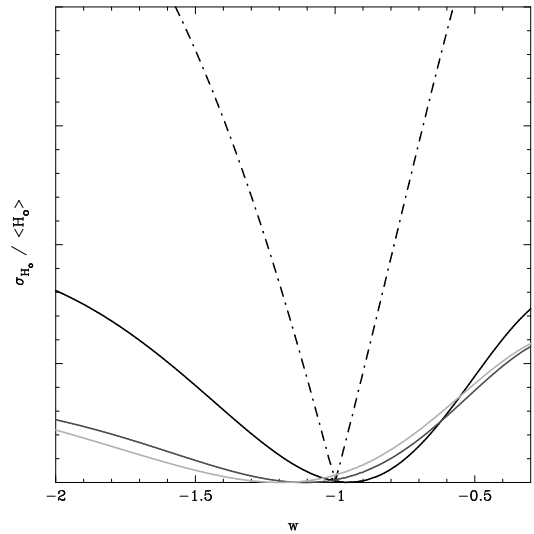} }
\caption{The  cosmological   minimization  function,  $\sigma_{H_o}  /
\left<H_o\right>$, versus  the quintessence equation of  state for the
chosen fiducial  cosmology and a  sample of one  hundred gravitational
lens  system. The  dot-dash line  represents  the ideal  case with  no
noise; as can be seen this  is zero at $w=-1$.  The black line assumes
an uncertainty in the time-delay and lens modeling of 5\%, the lighter
curve  of  10\%  and the  lightest  15\%.   A  DC component  has  been
subtracted from these latter curves.}
\label{fig4a}
\end{figure}

As seen from  Equation~\ref{hubbletime}, the cosmological component of
the measure  of Hubble's constant  is dependent upon a  combination of
angular diameter  distances. Table~\ref{table2} presents  the relative
values  in the  measured Hubble's  constant for  the  linear evolution
models    described     above,    for    the     lens    systems    in
Section~\ref{timedelay}.  The  largest changes  are seen in  the first
two rows  which represent fiducial  models; these possess  the highest
lens  redshifts in the  sample. For  the observed  lens system,  it is
clear   that  the   evolving  quintessence   models  do   differ  from
non-evolving  models,  this   difference  is  not  substantial,  being
typically less than a few percent for evolution models with $w_1=0.4$.

\begin{figure*}
\centerline{ \psfig{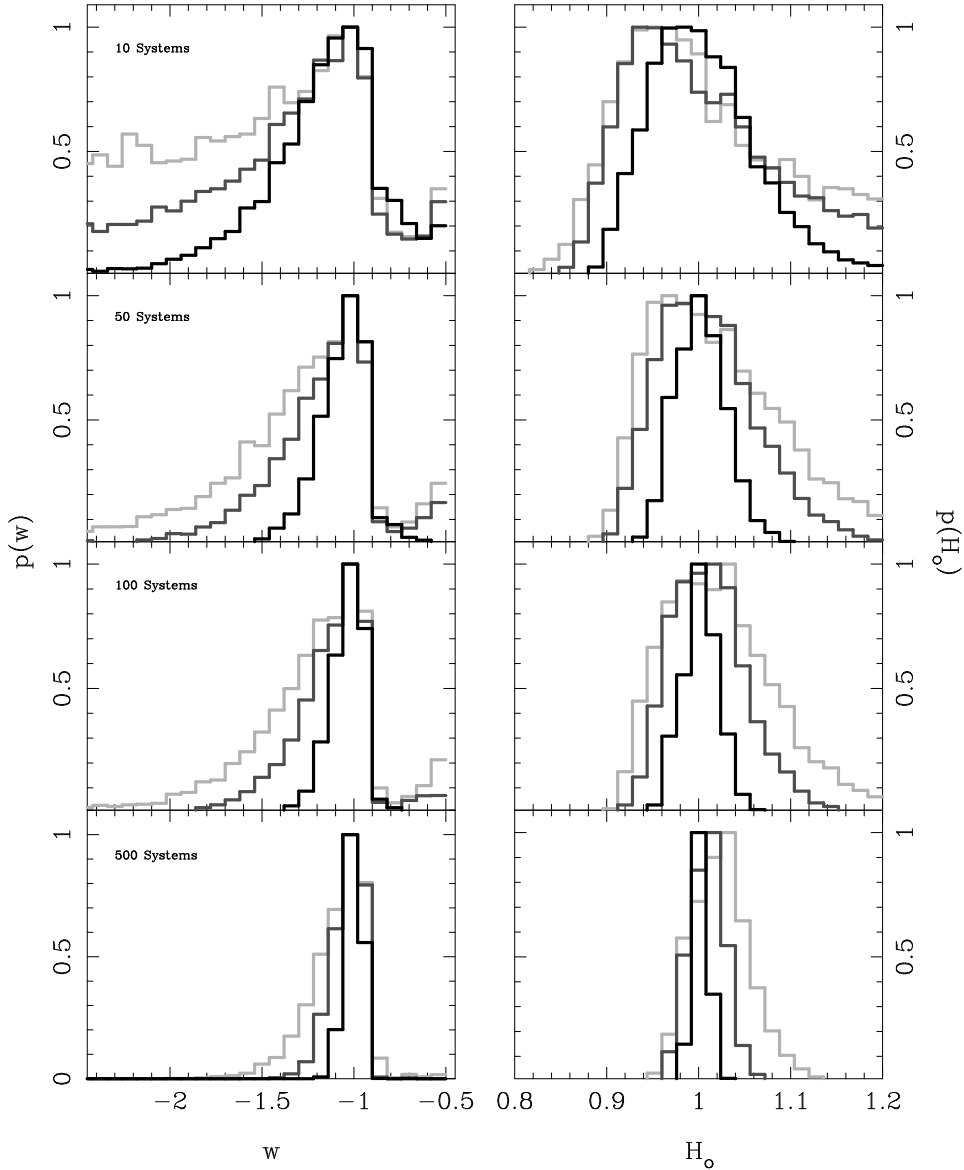} }
\caption{The   results  of   the  Monte   Carlo  simulations   of  the
determination  of  the quintessence  equation  of  state and  Hubble's
constant from  the measurement of  samples of gravitational  lens time
delays. The  black line assumes  an uncertainty in the  time-delay and
lens modeling of 5\%, the lighter curve of 10\% and the lightest 15\%}
\label{fig5}
\end{figure*}

\section{Monte Carlo Simulations}\label{montecarlo}
How many lensed  systems are required before firm  limits can be made?
To  address this  question a  number of  Monte Carlo  simulations were
undertaken with a population of lensed sources, assuming that for each
a time  delay could be determined  to a specified  accuracy.  For real
lensed  systems, additional  uncertainty is  introduced from  the lens
modeling.   Here,  however,  it  is  assumed that  the  lens  modeling
introduces  no systematic  uncertainty and  that the  random  error is
taken into account as an  additional source of uncertainty in the time
delay.

A  number  of  programs  have  examined  the  relative  statistics  of
gravitational  lensing  (e.g. Kochanek  1993a;  1993b; Williams  1998;
Falco et al. 1998; Helbig et  al. 1999). The complex analyses in these
studies go beyond this paper  and a more straight forward approach was
adopted;  this was  to examine  the  distribution of  source and  lens
redshift pairs in the CASTLES  database (Mu{\~ n}oz et al.~1998). This
is seen  to be a  scatter plot of  sources between redshifts 1  and 4,
with  lenses between  0.2 and  1.1.  Source $(z_s)$  and lens  $(z_l)$
redshifts  where, therefore,  selected  randomly in  these range,  but
ensuring  that $z_s  > 4z_l  - 2.2$.  This cut  ensures that  lens and
sources are always reasonably separated in redshift space.

In   choosing  a   background   cosmology,  a   fiducial  model   with
$\Omega_m=0.3$  and  $\Omega_w=0.7$  with  a $w=-1$  and  $H_o=1$  was
employed. In an ideal universe,  where there are no measurement errors
and where  gravitational lens models  can be uniquely  determined, the
analysis of  a sample of systems  may still produce  a distribution of
$H_o$  values,  as an  incorrect  choice  of cosmological  parameters,
including   the  quintessence  equation   of  state   $w$,  influences
differently  each  system (e.g.   Figure~\ref{fig1}).   In this  ideal
universe,  however,  all  one  would   need  to  do  is  to  vary  the
cosmological parameters  until all systems yielded the  same value for
the Hubble  constant. In this way,  not only would  $H_o$ be measured,
but the underlying cosmology would be determined as well.

In  the   real  universe,   however,  the  influence   of  measurement
uncertainties needs  to be considered.  Instead of  obtaining a unique
value of Hubble's constant, one could vary the cosmological parameters
such that the  dispersion in the resultant $H_o$'s  are minimized.  An
example of  this for  a population of  one hundred  gravitational lens
systems is given in Figure~\ref{fig4a}; note that the function that is
minimized is $\sigma_{H_o} / \left<H_o\right>$ as it is the fractional
dispersion of the results that  is of interest. The dot-dashed line is
the ideal  universe case, where  there are no sources  of uncertainty,
which  possesses a  minimum value  (of  zero) at  the chosen  fiducial
value.   The solid  curves represent  varying values  of noise  in the
determination of the time delay,  the black being 5\% noise, grey 10\%
and light  grey 15\%  noise.  Clearly this  function broadens  as more
noise is  added, and  the minimum is  not necessarily at  the fiducial
value of $w$.

\begin{table}
 \centering
 \begin{tabular}{ccccc}
  & 10 & 50 & 100 & 500 \\ \hline
%
%
%
 5\% & (-3.52,-0.14) & (-1.39,-0.78) & (-1.25,-0.89) & (-1.10,-0.92)\\
10\% & (-4.41,-0.13) & (-3.85,-0.15) & (-1.70,-0.54) & (-1.23,-0.92)\\
15\% & (-4.45,-0.13) & (-4.52,-0.13) & (-4.09,-0.14) & (-1.48,-0.88)\\ \hline
 5\% &  (0.91,1.30) &  (0.97,1.07) &  (0.98,1.04) &  (0.99,1.02)\\
10\% &  (0.88,1.39) &  (0.92,1.36) &  (0.95,1.13) &  (0.99,1.05)\\
15\% &  (0.87,1.42) &  (0.92,1.42) &  (0.93,1.39) &  (0.99,1.10)\\ \hline
\end{tabular}
\caption{\label{table3} The  95\% confidence intervals  for $w$ (upper
values) and $H_o$ (lower values) for the Monte Carlo samples described
in the text.  The sample sizes  are given across the top of the table,
and each row is for an  assumed error in the determination of the time
delay. Note  that in calculating these  confidence intervals, outlying
points with $w>0$ have been neglected so that they represent the width
of the distribution.}
\end{table}

Figure~\ref{fig5} presents  the results of  undertaking this procedure
for  various samples of  gravitational lens  systems.  The  left panel
presents the probability distribution for the quintessence equation of
state $w$ while the right hand panel is the corresponding distribution
in  Hubble's constants;  note  that all  the  distributions have  been
normalized to a peak value of one.  The top row is for a sample of ten
gravitational lens  systems, akin to the situation  today, followed by
50, 100 and 500 systems. Ten thousand realizations were undertaken for
each sample  size. Each  line on the  plot corresponds to  a different
level of uncertainty in the  gravitational lens time delay as outlined
in  the previous  paragraph.  Clearly  today,  where there  are but  a
handful of gravitational lens systems for which we have determined the
time  delay, with  overall  uncertainties exceeding  5-10\%, then  the
resultant  Hubble's constant  and $w$  that  can be  derived from  the
sample are unlikely to accurately represent the underlying values.

Table~\ref{table3}  presents  the  95\%  confidence interval  for  the
estimation of $w$  and $H_o$ for the various  samples.  Increasing the
sample  size greatly improves  the situation,  although even  with 500
lenses and  15\% noise, the values  of $w$ and $H_o$  are not strongly
constrained. One is led to conclude, therefore, that a large sample of
lenses with very accurately determined  time delays and lens models is
required  to significantly determine  the the  underlying cosmological
parameters.   Given the  observational effort  in such  a  task, other
approaches to probing the cosmological equation of state are likely to
prove more  fruitful.  Given  this, the analysis  was not  extended to
consider  the  smaller   influences  of  quintessence  evolution  (see
Section~\ref{evolve}).

\section{Do current lens systems suggest $w<-1$?}\label{speculate}
Can  the current  observations of  time delays  in  gravitational lens
systems  tell  us  anything  about   the  equation  of  state  of  the
quintessence component?   In recent  years, dedicated monitoring  of a
number of  lensed systems has provided accurate  $(\sim 10-20\%)$ time
delay  determinations  (e.g.    Fassnacht  et  al.~1999;  Koopmans  et
al. 2000). An  examination of the Hubble's constant  derived from such
studies reveals that, typically, it  is less than the value determined
from  local   studies,  even  accounting   for  standard  cosmological
differences (e.g.  Impey  et al. 1998; Koopmans et  al.  2001; Winn et
al. 2002). This very question  was also recently addressed by Kochanek
(2002) who  suggests  that  this  discrepancy is  potentially  due  to
galaxies  possessing concentrated  dark matter  halos with  a constant
mass-to-light ratios, at odds  with expectations from cold dark matter
structure models.  Here, an alternative solution is considered.

Examining    the    panel   in    Figure~1    corresponding   to    an
$\Omega_m=0.3,\Omega_w=0.7$  cosmology, it  is apparent  that choosing
$\Omega_m=1$  (equivalent  to setting  $w=0$)  results  in almost  the
lowest possible determination of $H_o$. Considering a cosmology with a
classical  cosmological  constant  ($w=-1$) increases  the  determined
value of  $H_o$ by  $\sim5-20\%$ (dependent upon  the source  and lens
redshifts), but as noted  above, the currently determined values still
tend to lie below the  72${\rm km/s/Mpc}$ derived locally (Freedman et
al.  2001).  Assuming that  the gravitational lens models are correct,
one  way to  reach  concordance  between the  two  approaches is  that
$w<-1$;  such a  conclusion is  consistent with  the  recently derived
limit  of  $w<-0.85$ from  an  analysis  of  a combination  of  cosmic
microwave background, high redshift supernovae, cluster abundances and
large  scale structure  data (Wang  et al.   2000; Bean  \& Melchiorri
2001).   While tantalizing,  however,  it must  be  conceded that  the
current   differences  in   the  approaches   are   not  statistically
significant, especially given the  relatively large uncertainty in the
modeling of  gravitational lens mass distributions  (see, for example,
the range of  $H_o$ values obtained from the  modeling of PG 1115+080;
Kochanek 2002).   Given a large sample of  gravitational lens systems,
however, the  determination of a systematic difference  in the derived
Hubble's constant could be made.

\section{Conclusions}\label{conclusions}
This paper  has investigated the  role of a quintessence  component on
angular  diameter  distances,  specifically  their  influence  on  the
determination  of Hubble's  constant from  the measurement  of  a time
delay in multiply imaged quasars.

For  flat universes,  with an  unevolving quintessence  component, its
seen that, for a gravitational lens system in which the time delay has
been measured,  the resultant Hubble's constant is  dependent upon the
value  of the  equation of  state parameter  $w$.   Interestingly, the
dependence  of the  determined value  of  the Hubble's  constant as  a
function of $w$  possesses a minimum which is  independent on the lens
and source redshift.

Several models of evolving quintessence were also examined, consisting
of  a linear evolution  of the  equation of  state with  redshift. The
cosmologies resulted  in significantly different forms  of the angular
diameter distance. Hence, our view of the cosmos would be different in
the various cosmologies. When  considering the specific combination of
angular   diameter   distances   that  constitute   the   cosmological
contribution  to the gravitational  lensing determination  of Hubble's
constant, it is seen that the resulting variations between cosmologies
is  very  small, a  matter  of  only a  few  percent,  relative to  an
unevolving case with the same present day constitution.

A number of  Monte Carlo simulations of the  determination of Hubble's
constant and the quintessence  equation of state, $w$, were undertaken
to explore  the efficacy  of this approach.   These revealed  that the
present situation with only a handful of lensed systems does not allow
an accurate determination of the cosmic equation of state, and that at
least an order of magnitude  more lenses are truly required to provide
a reasonably  robust determination  of the underlying  cosmology.  The
next generation of all-sky surveys are presently underway (e.g.  Sloan
Digital Sky  Survey) or  are being planned  (e.g.  VISTA,  PRIME), and
these  datasets will  greatly increase  the number  of  lensed quasars
available for  monitoring studies.  Cooray \&  Huterer (1999) estimate
that  $\sim 2000$  lensed quasars  will  be identified  from the  SDSS
database  alone, and a  much larger  number can  be expected  from the
deeper VISTA and  PRIME surveys. The number of  these sources amenable
for follow-up monitoring campaigns will naturally be much smaller, but
one  can confidently  expect a  sample of  several hundred  systems to
eventually become  available.  However,  given the effort  required to
first find such systems, as well as monitor them to determine the time
delays and the modeling procedure, it is likely that $w$ will be first
determined using  one of the other various  techniques currently being
proposed.  We conclude, therefore, that gravitational lens time delays
are likely to prove poor probes of the universal equation of state.

\newfont{\afont}{cmfi10}
\section*{Acknowledgements}
The anonymous referee is thanked for comments that improved the paper.
GFL acknowledges  using David W. Hogg's  wonderful ``Distance measures
in cosmology''  cheat sheet  ({\tt astro-ph/9905116}), and  thanks the
{\afont  Gorillaz}  for their  self-titled  album.   Terry Bridges  is
thanked for providing the computational cycles on {\bf \tt Odin}.

\newcommand{\aap}{A\&A}
\newcommand{\apj}{ApJ}
\newcommand{\apjl}{ApJ}
\newcommand{\aj}{AJ}
\newcommand{\mnras}{MNRAS}
\newcommand{\apss}{Ap\&SS}
\newcommand{\nat}{Nature}

\end{document}